\documentclass[epj]{svjour}
\usepackage{graphicx}
\usepackage{amsmath}
\usepackage{subfig}
\usepackage{overpic}
\usepackage{braket}

\newcommand{\abs}[1]{\ensuremath{\left\vert#1\right\vert}}

\title{Partitioning Schemes and Non-Integer Box Sizes for the Box-Counting Algorithm in Multifractal Analysis}
\titlerunning{Partitioning Schemes for the Box-Counting Algorithm in Multifractal Analysis}
\author{Stefanie Thiem \and Michael Schreiber}
\institute{Institut f\"ur Physik, Technische Universit\"at Chemnitz, D-09107 Chemnitz, Germany}

\abstract{
We compare different partitioning schemes for the box-counting algorithm in the multifractal analysis by computing the singularity spectrum and the distribution of the box probabilities. As model system we use the Anderson model of localization in two and three dimensions. We show that a partitioning scheme which includes unrestricted values of the box size and an average over all box origins leads to smaller error bounds than the standard method using only integer ratios of the linear system size and the box size which was found by Rodriguez et al.\ \cite{EPJB.2009.Rodriguez} to yield the most reliable results.
}

\begin{document}
\maketitle

\vspace{0.3cm}{\large\scshape\hrule\vspace{0.1\baselineskip}}\vspace{0.3cm}

\section{Introduction}

Systems with multifractal properties are often found in nature, e.g. in biological systems, in meteorology and geology, or in economics, e.g. stock market time series. For a detailed study of these properties multifractal analysis plays an important role. Also the electronic wave functions $\Phi$ of physical models like the Anderson model of localization at the mobility edge, quasiperiodic tilings and chaotic systems show multifractal properties \cite{JPhysA.1986.Castellani,PhysRevLett.1991.Schreiber,PhysStatSol.1990.Boettger,PhysRevLett.1989.Chabra}. One can distinguish two methods for the multifractal analysis: the system-size scaling approach and the box-size scaling approach \cite{PhysRevB.2008.Rodriguez3}. In both approaches the system is partitioned into boxes and one measures the probability to find the electron in a particular box, i.e., we study the spatial distribution of the wave functions \cite{PhysRevB.1997.Huckestein}. While for the system-size scaling the box size is kept constant and the system size is varied, for box-size scaling one studies how the probability distribution changes with the box size. The latter method is more common because it requires less computing time and storage.

For box-size scaling different partitioning schemes are known, where the scheme has a crucial influence on the fit results and the statistical analysis. Rodriguez et al. compared several approaches for partitioning the system into boxes including different box shapes and box sizes as well as an averaging over different box origins \cite{EPJB.2009.Rodriguez,PhysRevB.2008.Rodriguez3}. They found that a partitioning of the system into cubic boxes with integer ratios of the linear system size $N$ and the linear box size $L$ is numerically most reliable in comparison to the other studied methods. However, a previous work by one of us showed that the fit quality can be improved by adapting the multifractal analysis to work also for non-integer ratios $N/L$ and by averaging over different origins of the boxes \cite{PhysRevLett.1991.Schreiber,ChemPhys.1993.Grussbach,Book.1996.Hoffmann}. Unfortunately, this method was not considered in the comparative study by Rodriguez et al. Therefore, we compare in this paper the fit quality of the best method proposed in \cite{EPJB.2009.Rodriguez} with the participation scheme allowing non-integer ratios $N/L$ \cite{PhysRevLett.1991.Schreiber}. We show that the latter partitioning scheme yields smaller errors in the least squares fits than the standard scheme which allows only integer ratios of $N/L$.

The paper is structured in the following way: First, we briefly introduce the Anderson model in Sec.\ \ref{sec:anderson} and the multifractal analysis in Sec.\ \ref{sec:multifrac}. We then discuss different partitioning schemes for the box-counting algorithm in Sec.\ \ref{sec:partitioning} and compare their performances in Sec.\ \ref{sec:results}. This is followed by a brief conclusion in Sec.\ \ref{sec:conclusion}.

\section{Anderson Model of Localization}
\label{sec:anderson}

\begin{figure}
  \centering
  \includegraphics[width=\columnwidth]{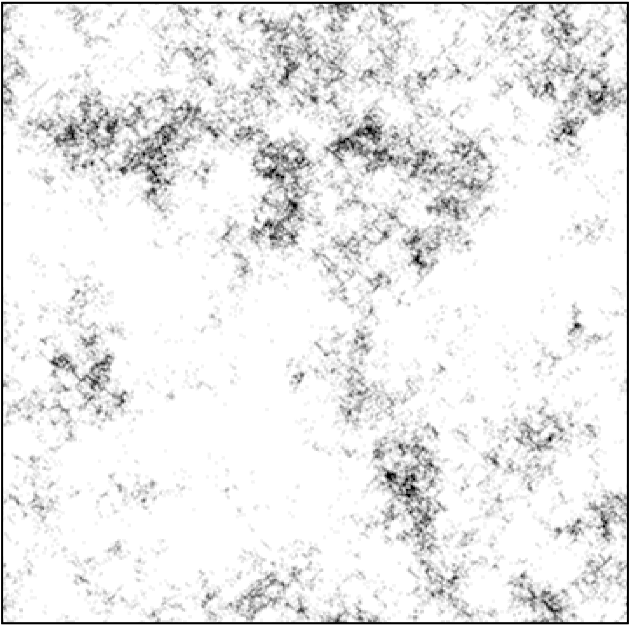}\\
  \caption{Probability amplitude $\abs{\Phi_\mathbf{r}}$ of a wave function on a square lattice of $N^2 = 360 \times 360$ sites. Every site $\mathbf{r}$ with an amplitude larger than the average is shown, i.e., $\abs{\Phi_\mathbf{r}} > 1/N$ and four different grey levels distinguish whether $\abs{\Phi_\mathbf{r}}^2 > 2^j/N^2$ for $j=0, 1,2,3$.}
  \label{fig:wavefunction}
\end{figure}

As model system for studying the influence of the partitioning in the multifractal analysis we use the eigenstates of the tight-binding Anderson Hamiltonian
\begin{equation}
 \label{equ:anderson}
 \mathbf{H} = \sum_{\mathbf{r},\mathbf{r}^\prime}^\mathrm{n.n.} \ket{\mathbf{r}} \bra{\mathbf{r}^\prime} + \sum_\mathbf{r} \epsilon_\mathbf{r} \ket{\mathbf{r}} \bra{\mathbf{r}} \;,
\end{equation}
represented in the orthogonal basis of states $\ket{\mathbf{r}}$ associated to a vertex $\mathbf{r}$. The onsite potential $\epsilon_\mathbf{r}$ is chosen according to a uniform distribution in the interval $[-w/2,w/2]$ for a given disorder strength $w$. Each site $\mathbf{r}$ is connected to the nearest neighbors in a regular square/cubic lattice with periodic boundary conditions. For a given realization of disorder we consider only the eigenstate $\Phi_\mathbf{r}$ of the Anderson Hamiltonian closest to the band center at $E=0$. The wave functions are obtained using the program JADAMILU, which is based on the Jacobi-Davidson method \cite{CompPhysCom.2007.Bollhoefer} and allows one to compute selected eigenvalues and eigenfunctions of large sparse matrices with high efficiency. In Fig.\ \ref{fig:wavefunction} we visualize a two-dimensional wave function for the disorder strength $w = 4$ for the linear system size $N =360$.

The investigations in this paper are restricted to moderate system sizes with at most $N =360$ in two dimensions and $N = 60$ in three dimensions which are sufficiently large for the presented purpose. We have also studied larger system sizes up to $N = 720$ in two dimensions and up to $N = 90$ in three dimensions but we have not found any significant differences for the fit results for the single wave functions. Of course, the system size has a crucial influence on the ensemble average, and the below discussed measures are not yet completely size-independent for these system sizes. Therefore, large system sizes should be preferred. As we are only interested in the optimization of the partitioning scheme we mainly consider single wave functions.

\section{Multifractal Analysis}
\label{sec:multifrac}

The multifractal analysis is based on a standard box-counting algorithm, i.e., in $d$ dimensions the system of size $V = N^d$ is divided into $B = N^d / L^d$ boxes of linear size $L$ \cite{JPhysA.1986.Castellani,PhysRevLett.1989.Chabra,PhysRevA.1986.Halsey,Book.1988.Feder}. For each box $b$ we measure the probability to find the electron in this particular box. Hence, for an eigenstate $\Phi_\mathbf{r}$ and the box size $L$ we determine the probability for the $b$th box \cite{PhysRevLett.1989.Chabra}
\begin{equation}
    \mu_b(\Phi, L) = \sum_{\mathbf{r} \in \,\mathrm{box }\, b} \abs{\Phi_\mathbf{r}}^2
\end{equation}
and construct for each parameter $q$ a measure by normalizing the $q$th moments of this probability
\begin{equation}
    \label{equ:probmu}
    \mu_b(q, \Phi, L) = \frac{ \mu_b^q(\Phi, L)} { P(q,\Phi,L)  } \;,
\end{equation}
where
\begin{equation}
P(q,\Phi,L) = \sum_{b=1}^{B} \mu_b^q(\Phi, L) \;.
\end{equation}
The mass exponent of a wave function
\begin{equation}
    \label{equ:tau}
    \tau_q(\Phi) = \lim_{\varepsilon \to 0} \frac{ \ln P(q,\Phi,L)} {\ln \varepsilon} 
\end{equation}
can be obtained by a least squares fit of $\ln P$ versus $\ln \varepsilon$, where $\varepsilon = L/N$. The generalized dimensions are easily obtained as $D_q = \tau_q / (q-1)$.

According to Eq.\ \eqref{equ:tau} small boxes should be considered for the least squares fit. Further, the power-law behavior defined in Eq.\ \eqref{equ:tau} requires the absence of a typical length scale in the distribution of the probability density of the wave functions. Therefore, it should be computed for box sizes $l \ll L \ll N$ to properly display the multifractal properties, where $l$ corresponds to microscopic length scales of the system like e.g.\ the lattice spacing \cite{IntJModPhys.1994.Janssen}. Rodriguez et al.\ used box sizes from $L = 10$ to $N/2$ \cite{EPJB.2009.Rodriguez}. For reasons of comparison we have performed most of the calculations also for this range, but we also show that the range of considered box sizes has a crucial influence on the observed results.

Another quantity often used for the multifractal analysis is the singularity spectrum $f(\alpha_q)$, which denotes the fractal dimension of the set of points where the wave-function intensity behaves according to $\abs{\Phi_\mathbf{r}}^2 \propto N^{-\alpha_q}$, which means that in a discrete system the number of such points scales as $N^{f(\alpha_q)}$ \cite{PhysRevB.2008.Vasquez}. The singularity spectrum $f(\alpha_q)$ follows from a parametric representation of the singularity strength 
\begin{equation}
    \alpha_q(\Phi) = \lim_{\varepsilon \to 0} \frac{A(q,\Phi, L) }{\ln \varepsilon}
\end{equation}
with
\begin{equation}
    \label{equ:A}
    A(q,\Phi, L) = \sum_{b=1}^{B} \mu_b(q, \Phi, L) \ln \mu_b(1, \Phi, L)
\end{equation}
and the fractal dimension
\begin{equation}
    f_q(\Phi) = \lim_{\varepsilon \to 0} \frac{F(q, \Phi, L) }{\ln \varepsilon}
\end{equation}
with
\begin{equation}
    \label{equ:F}
    F(q,\Phi, L) = \sum_{b=1}^{B} \mu_b(q, \Phi,L) \ln \mu_b(q, \Phi, L) \;.
\end{equation}
Both quantities can be determined by least squares fits of $A(q,\Phi,L)$ or $F(q,\Phi,L)$ versus $\ln \varepsilon$ considering different box sizes $L$. The mass exponents $\tau_q$ and the generalized dimensions $D_q$ can also be obtained from the singularity spectrum $f(\alpha_q)$ via the Legendre transformation $ \tau_q = D_q(q-1) = q \alpha_q  - f(\alpha_q)$ \cite{PhysRevA.1986.Halsey}.

Regarding the partitioning of the system into boxes several options are possible, which are discussed in more detail in the following two sections.

\section{Partitioning Schemes}
\label{sec:partitioning}

In this section we briefly introduce the considered methods for the optimization of the partitioning in multifractal analysis. In particular, we discuss the following three partitioning schemes for the box-counting algorithm:
\begin{enumerate}
 \item[A)] cubic boxes with integer ratios $N/L$,
 \item[B)] cubic boxes with unrestricted values for $L$, and
 \item[C)] cubic boxes with integer ratios $N/L$ including an average over all box origins.
\end{enumerate}
To be specific, we define the left upper corners of a box as box origin in the figures below.

\mathversion{bold}
\subsection{Scheme A: Cubic boxes with integer ratios $N/L$}
\mathversion{normal}

\begin{figure}[b!]
  \centering
  \includegraphics[width=0.49\columnwidth]{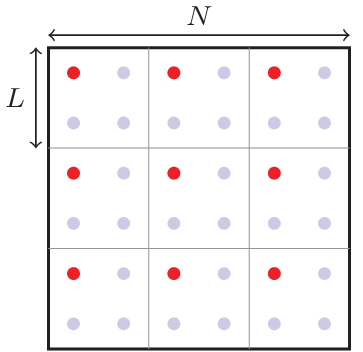}
  \includegraphics[width=0.49\columnwidth]{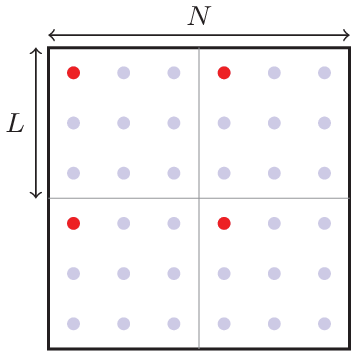}
  \caption{Partitioning scheme with integer ratio $N/L$ for a system of linear size $N = 6$ and box sizes $L=2$ (left) and $L=3$ (right). The red dots indicate the considered box origins.}
  \label{fig:integer-ratios}
\end{figure}

This quite simple partitioning scheme is the most straightforward procedure in which only such box sizes are considered for which the boxes cover the system gapless without overlap as visualized in Fig.\ \ref{fig:integer-ratios}. Hence, the ratio $1/\varepsilon = N/L$ is restricted to integer values. Obviously this condition is a severe limitation of the possible choices for $N$ and $L$. For example, in the case of $N = 6$ the two partitionings with $L = 2$ and $L=3$ shown in Fig.\ \ref{fig:integer-ratios} are the only possibilities. For $N = 5$ or $N = 7$ there does not exist any such way of subdividing the system.

Exploiting periodic boundary conditions, one could utilize different box origins. But usually no additional average over different box origins is considered.
Rodriguez et al. studied several partitioning schemes and found that this one shows the most reliable performance of the considered schemes \cite{EPJB.2009.Rodriguez}. However, we show in this paper that a method considering unrestricted values of box sizes shows a better performance. This scheme was already successfully applied for the multifractal analysis \cite{PhysRevLett.1991.Schreiber,ChemPhys.1993.Grussbach}. Unfortunately, it was not taken into account in the comparison by Rodriguez et al. \cite{EPJB.2009.Rodriguez}.

\mathversion{bold}
\subsection{Scheme B: Cubic boxes with unrestricted values for $L$}
\mathversion{normal}

\begin{figure}[b!]
  \centering
  \includegraphics[width=\columnwidth]{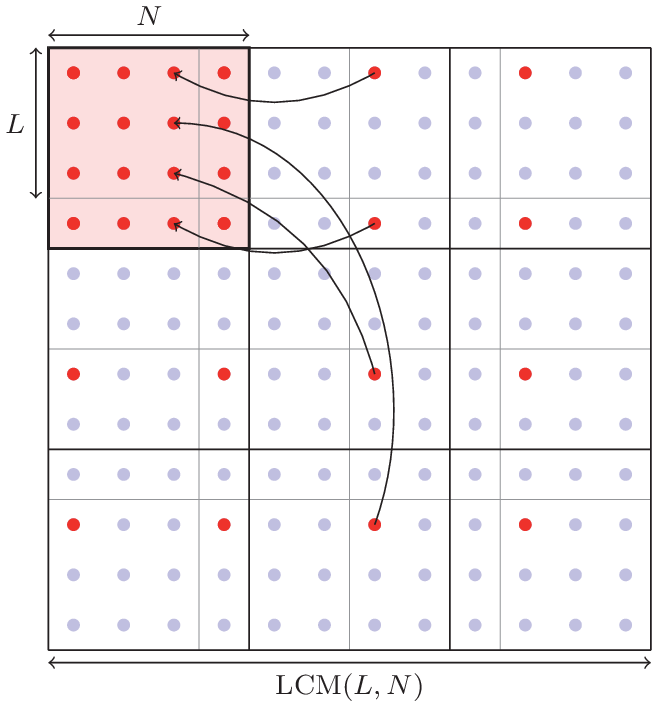}
  \caption{Partitioning scheme for unrestricted values of the box size $L$ for a system of linear size $N = 4$ and the box size $L = 3$. The original system is repeated LCM$(L,N)/N = 3$ times in each direction so that the resulting system of linear size LCM$(L,N) = 12$ can be filled gapless by the boxes. For coprime $L$ and $N$ this method intrinsically includes an averaging over all box origins as indicated by the red box origins. The arrows visualize this for the sites in the third column of the original system.}
  \label{fig:all-ratios}
\end{figure}

The fitting of the scaling exponents in the box-size scaling can be enhanced if all possible box sizes $L$ as well as an average over all box origins is considered \cite{PhysRevLett.1991.Schreiber,ChemPhys.1993.Grussbach,Book.1996.Hoffmann}. This method has a rather simple geometrical interpretation, which is displayed in Fig.\ \ref{fig:all-ratios}. First, the original system is periodically repeated LCM$(L, N)/N$ times (LCM labels the least common multiple) in each direction. The new system can then be covered gapless for any linear box size $L$. If $N$ and $L$ are coprime, then this method automatically includes an average over all different box origins as shown in Fig.\ \ref{fig:all-ratios}. For the other cases we explicitly consider an average over all different box origins.

From the computational perspective it is easier not to look at the periodically repeated system but rather to imagine that all boxes with origins outside the original system are folded back so that their origins fall into the original system. One can easily see from Fig.\ \ref{fig:all-ratios} that if $N$ and $L$ are not coprime such a folding back would not yield all sites of the original system, but it is of course straightforward to generalize the idea and to place a box of linear size $L$ on each site of the original system in this case, too.
Hence, in all cases each of the $N^2$ sites is considered $L^2$ times and the proper average for a quantity $X$ is obtained by
\begin{equation}
    \label{equ:average}
    \langle X \rangle  = \frac{1}{L^2} \sum\limits_\mathrm{box \; origins} X \;.
\end{equation}
This average then replaces the simple sum over all $B$ boxes.

Rodriguez et al. also studied a method which allows one to utilize all box sizes $L$ \cite{EPJB.2009.Rodriguez}. In contrast to our method, they did not periodically repeat the system LCM$(L, N)/N$ times. Hence, they faced the problem that different effective system sizes occur, and they had to introduce an additional average over these different effective system sizes. Our method does not require such an additional averaging because the boxes fill the new system gapless, so that no effective system sizes appear.

\mathversion{bold}
\subsection{Scheme C: Cubic boxes with integer ratios $N/L$ including an average over all box origins}
\mathversion{normal}

This method is a mixture of the two former partitioning schemes. Now we only consider integer ratios but include the average over all box origins as in the previous method. Rodriguez et al. also studied the average over all initial box origins but pointed out that for an ensemble average over many wave functions this method reduces the standard deviation almost to zero \cite{EPJB.2009.Rodriguez}. The reason for this apparently negative side effect is that they counted each box origin as a different disorder configuration, which is clearly not the case.

Hence, a proper ensemble average over different disorder configurations can still be obtained by computing the average quantities $P(q,\Phi,L)$, $A(q,\Phi,L)$ and $F(q,\Phi,L)$ for all box origins for each wave function according to Eq.\ \eqref{equ:average} but without considering each box origin as a new disorder configuration. In this case no unwanted reduction of the standard deviation of the ensemble average occurs. However, the average over the box origins in Eq.\ \eqref{equ:average} means that the data for each wave functions are evaluated with a better accuracy, so that a reasonable reduction of the standard deviation of the ensemble average can be expected.

To avoid a misunderstanding, we point out that the replicas in Fig.\ \ref{fig:all-ratios} are used for an easy explanation of our scheme only. They are not necessary for our computation. One can also utilize the following interpretation: Due to periodic boundary conditions, no site of the system can or should be distinguished or preferred over any other site. But the choice of the system origin makes such a distinction. This bias can be avoided by treating every other site equivalently, by considering every site as a possible system origin. Then the generic placement of boxes starting at all the possible system origins is equivalent to our above placement of box origins on each site of the system.

\section{Comparison of Results}
\label{sec:results}

For a comparison of the three partitioning schemes we study the singularity spectrum of wave functions closest to the band center $E=0$ for certain disorder configurations. We consider wave functions in different dimensions and study also the influence of the considered range of box sizes.

\subsection{Two-dimensional wave functions}

\begin{figure}[b!]
    \centering
    \includegraphics[width=\columnwidth]{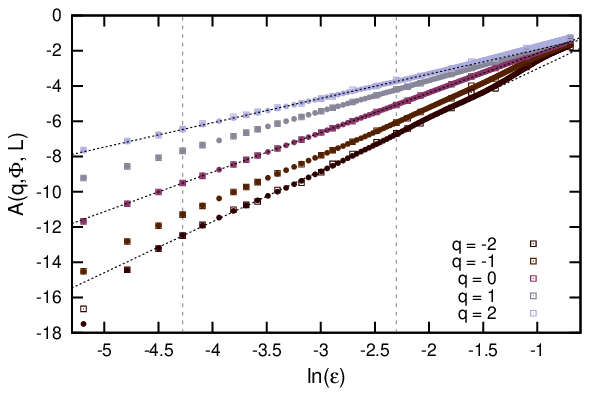}
    \includegraphics[width=\columnwidth]{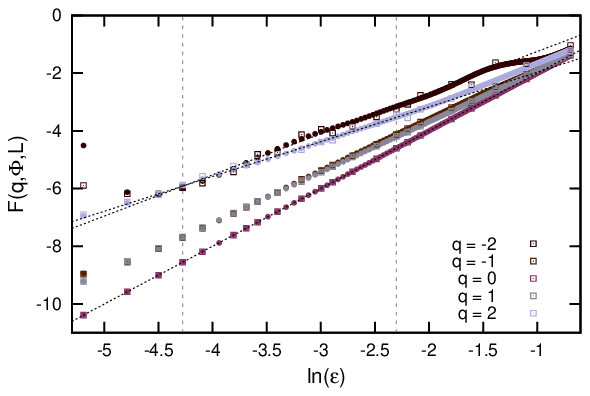}
    \caption{Scaling behavior of the quantities $A(q,\Phi,L)$ (top) and $F(q,\Phi,L)$ (bottom) with the ratio $\varepsilon = L/N$ of the box size $L$ and the linear system size $N$ for the two-dimensional wave function of the Anderson Hamiltonian shown in Fig.\ \ref{fig:wavefunction} with $N=360$ and disorder strength $w = 4$. The results are shown for the scheme A  with integer ratios $N/L$ (squares) and scheme B with unrestricted values of $L$ and a box-origin average (circles) for $2 \le L \le N/2$. The dotted lines correspond to least squares fits considering all box sizes in the range $5 \le L \le N/10$ marked by the two vertical dashed lines.}
    \label{fig:results-360-2D}
\end{figure}

\begin{figure}[t!]
    \centering
    \includegraphics[width=\columnwidth]{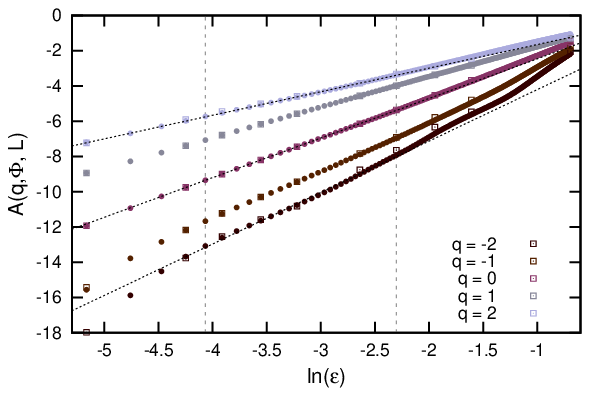}
    \includegraphics[width=\columnwidth]{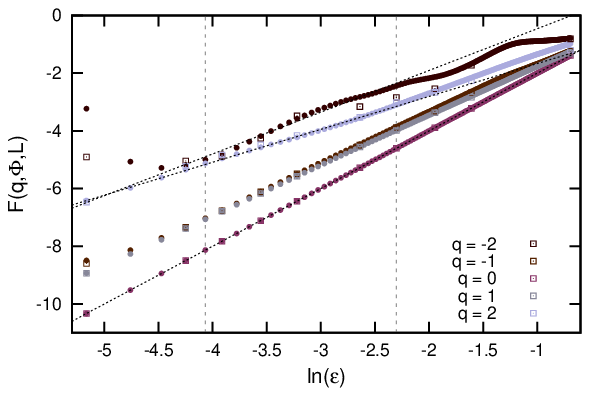}
    \caption{Same as Fig.\ \ref{fig:results-360-2D}, but for the system size $N = 350$.}
    \label{fig:results-350-2D}
\end{figure}

Figures \ref{fig:results-360-2D} and \ref{fig:results-350-2D} show the scaling behavior of the quantities $A(q,\Phi,L)$ and $F(q,\Phi,L)$ with the box size $L$ for a two-dimensional wave function for the system sizes $N = 350$ and $N = 360$. The results can be well described by a power law. The stronger fluctuations for negative values of $q$ are well known and can be explained by the less accurate values of the wave functions at vertices with small amplitudes, which are enhanced for negative powers of $\mu_b$ in Eq.\ \eqref{equ:probmu}. Apart from fluctuations for very small box sizes and negative $q$, the data are sufficiently close to straight lines as visualized in Figs.\ \ref{fig:results-360-2D} and \ref{fig:results-350-2D} so that we can determine the singularity strengths $\alpha_q(\Phi)$ and the fractal dimensions $f_q(\Phi)$ from the slopes. If not stated differently we consider box sizes in the range from $L = 10$ to $N/2$ for the computation of the singularity spectrum in accordance to the setup used by Rodriguez et al. \cite{EPJB.2009.Rodriguez,PhysRevB.2008.Vasquez}.

\begin{figure}[t!]
    \centering
    \includegraphics[width=\columnwidth]{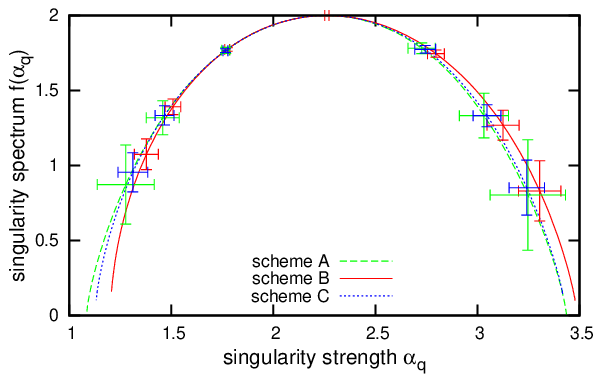}
    \includegraphics[width=\columnwidth]{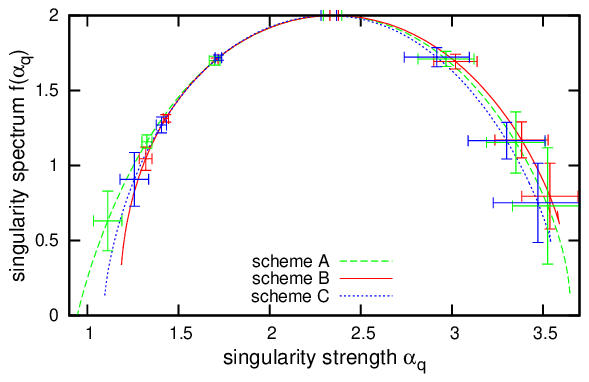}
    \caption{Singularity spectrum $f(\alpha_q)$ for the three different methods for the systems shown in Fig.\ \ref{fig:results-360-2D} with $N = 360$ (top) and in Fig. \ref{fig:results-350-2D} with $N = 350$ (bottom) with considered box sizes from $L = 10$ to $N/2$. The plot shows 1$\sigma$ error bars for integer $q$ from $-3$ to $3$. }
    \label{fig:singspec-2D}
\end{figure}

The corresponding singularity spectrum $f(\alpha_q)$ is shown in Fig.\ \ref{fig:singspec-2D}. From literature it is known that $f(\alpha_q)$ is a convex function of $\alpha_q$, where the maximum of $f(\alpha_q)$ corresponds to $q = 0$ with $f(\alpha_0) = d$ being the support of the measure \cite{PhysRevA.1986.Halsey,Book.1988.Feder}. This is verified in Fig.\ \ref{fig:singspec-2D} with $f(\alpha_0) \approx 2$. Further, for $q=1$ we have $f(\alpha_1) = \alpha_1$ and the limits $q \to \pm \infty$ yield the minimal/maximal value of $\alpha_q$, which corresponds to the singularity associated to the box containing the largest/smallest measure \cite{PhysRevLett.1991.Schreiber}.
For the periodic system, i.e.\ for vanishing disorder, the wave functions are Bloch waves extending over the entire system and therefore the singularity spectrum is expected to converge to the point $f(d) = d$. In the limit of a strongly disordered system with localized wave functions, the spectrum converges to the points $f(0) = 0$ and $f(\infty) = d$. The wave functions of the two-dimensional Anderson Hamiltonian show large spatial fluctuations, which increase with increasing disorder strength \cite{PhilMag.1992.Schreiber}. However, for moderate disorder strengths the wave functions can be approximately described by the multifractal analysis.

\begin{figure}[t!]
    \centering
    \includegraphics[width=\columnwidth]{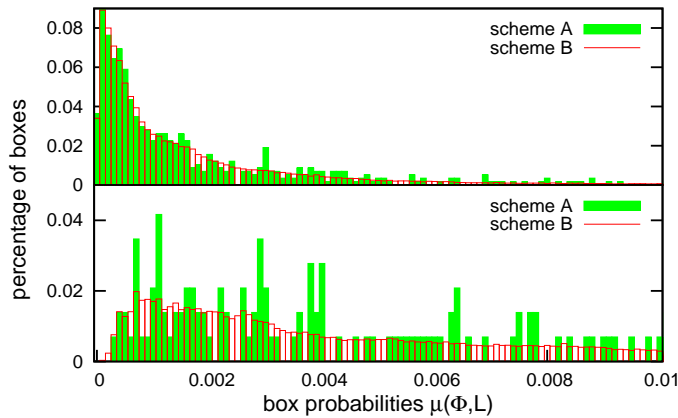}
    \caption{Distribution of box probabilities for the scheme A and scheme B for the wave function considered in Fig.\ \ref{fig:results-360-2D} with linear box sizes $L=15$ (top) and $L=30$ (bottom).}
    \label{fig:histogram}
\end{figure}

Comparing the results for the three different partitioning schemes we observe that the least squares fit of the quantities $A(q,\Phi,L)$ and $F(q,\Phi,L)$ includes many more points for the scheme B as for the other two methods, which makes the fitting much more reliable for scheme B. For the other two schemes the actual number of considered data points of course strongly depends on the number of factors of the system size $N$. Further, schemes A and B yield different values for the quantities $A(q,\Phi,L)$ and $F(q,\Phi,L)$ because scheme B considers all possible box origins in contrast to scheme A. The biggest differences are observed for large box sizes $L$ because in this case only very few box origins are taken into account in scheme A as visualized in Fig.\ \ref{fig:integer-ratios}.
The reason is that the number of considered box origins has a crucial influence on the distribution of the box probabilities $\mu_b(\Phi,L)$. This becomes clearly visible by plotting the distribution with and without an explicit average in Fig.\ \ref{fig:histogram}. While the distributions for small box sizes $L$ are quite similar, there are huge differences already for medium-sized boxes. Even for the small box sizes the distribution is not as smooth as the results averaged over all box origins. As we apply periodic boundaries there is no preferable choice for the box origins. Hence, the average over all box origins is a better estimate for the real box probability distribution. For the method C, which also includes an average over all box origins, the data points are identical to those of the partitioning scheme B but only box sizes for integer values of $1/\varepsilon$ are taken into account. Hence, the results for method B and method C are often rather close, but the accuracy of the fit for C is not as good as for B.

Having a look at the singularity spectra $f(\alpha_q)$ we observe that the results are rather close for all three schemes. However, the error bars are significantly larger for scheme $A$ than for scheme $B$ especially for the system size $N=360$. This is not surprising because in B many more data points are considered for the least squares fit and the average over all box origins leads to a better estimate of the box probabilities $\mu_b$. Comparing schemes B and C we observe that the errors are rather similar, i.e., most of the improvement of the fits is due to the averaging over all different box origins.
For the system with $N = 350$ the error bars in Fig.\ \ref{fig:singspec-2D} are rather close for all three schemes especially for negative values of $q$ and only a slight advantage of scheme B can be found. At first view this seems surprising because for this system even fewer box sizes are taken into account for scheme A and C. The reason is that by considering box sizes from $L = 10$ to $N/2$ in scheme B large boxes are strongly overrepresented in the least squares fits. This leads to a strong contribution of the non-linear behavior of $A(q,\Phi,L)$ and $F(q,\Phi,L)$ for large $L$ and negative $q$ to the fit in B (cp.\ Fig.\ \ref{fig:results-350-2D}) while for A and C almost no data points are considered in this range.

\mathversion{bold}
\subsection{Influence of range of box sizes $L$}
\mathversion{normal}

\begin{figure}[t!]
    \centering
    \hspace{0.2cm}\includegraphics[width=\columnwidth]{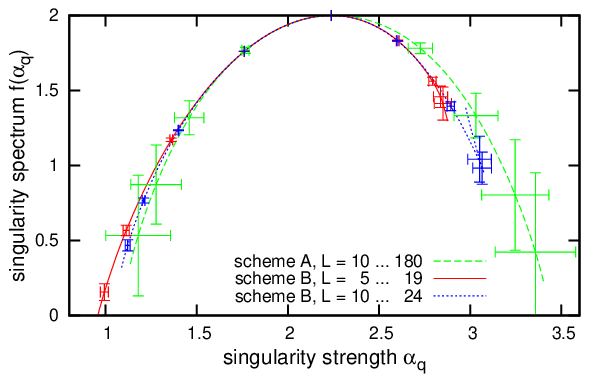}
    \includegraphics[width=0.96\columnwidth]{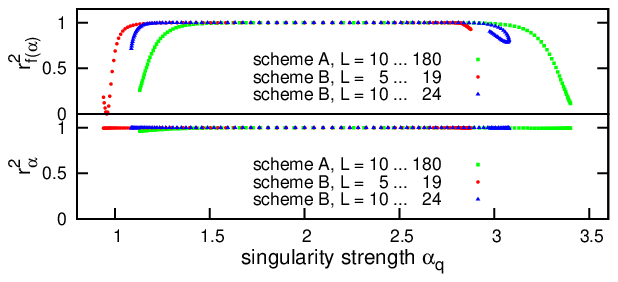}
    \includegraphics[width=0.96\columnwidth]{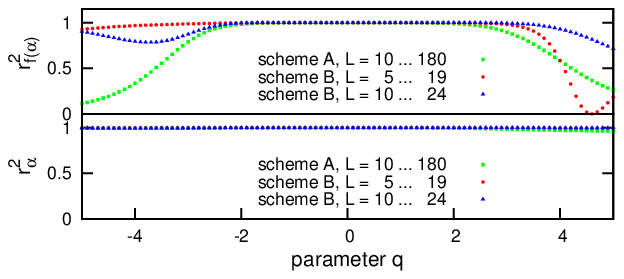}
    \caption{Singularity spectrum $f(\alpha_q)$ obtained using scheme A or scheme B with different ranges of box sizes $L$ for the wave function in Fig.\ \ref{fig:results-360-2D} with $-5 \le q \le 5$ and $1\sigma$ error bars for integer $q$ from $-4$ to $4$ (top). Corresponding squares $r^2$ of the correlation coefficients for the singularity strength $\alpha_q$ and the fractal dimension $f(\alpha_q)$ (center and bottom).}
    \label{fig:singspec-360-2Db}
\end{figure}

Further, we also investigated the influence of the considered range of box sizes. As mentioned in Sec.\ \ref{sec:multifrac}, the exponents should be computed from the scaling behavior of preferably small boxes and by considering the range $1 \ll L \ll N$ to properly measure the multifractal properties. The first point is not properly considered yet because especially in scheme B larger box sizes $L$ dominate the fit as mentioned above. In order to neglect the influence due to the different number of considered box sizes for the fits, we study the singularity spectrum for an identical number of box sizes for the different schemes. This means that we compute the singularity spectrum for the two-dimensional wave function in Fig.\ \ref{fig:wavefunction} by using exactly 15 different box sizes, i.e., for scheme A we have taken all box sizes with integer ratios $N/L$ in the range from $L = 10$ to $N/2$ into account and for scheme B we used on the one hand $L = 10$ to $24$ and on the other hand $L = 5$ to $19$. In the latter case we do not start with $L_\mathrm{min} = 2$ to assure $L \gg 1$ and, therefore, to omit the deviations from the power-law behavior which can be observed in Figs.\ \ref{fig:results-360-2D} and \ref{fig:results-350-2D} for small box sizes and negative values of $q$. This deviations are most likely caused by numerical inaccuracies which are enhanced for negative powers of $\mu_b$ and by summing over many boxes. The results in Fig.\ \ref{fig:singspec-360-2Db} again show that the error bars are significantly smaller for scheme B. We can also observe clear differences in the singularity spectrum especially for negative values of $q$ for all three considered setups.

We also plot in Fig.\ \ref{fig:singspec-360-2Db} the squares $r^2$ of the correlation coefficients obtained from the least squares fit, which is a measure for the amount of variability in the data that is accounted for by the fitting curve. Values of $r^2 = 1$ mean that the data can be fitted perfectly by a straight line, i.e., in this case we can identify a power-law behavior. While the numerical results in Fig.\ \ref{fig:singspec-360-2Db} show that for the singularity strength $\alpha_q$ the quantity $r^2 \approx 1$ for $-5 \le q \le 5$, we observe smaller values of $r^2$ for large positive and negative $q$ for the least squares fits of the fractal dimension $f_q$ in agreement with the increase of the fit errors for these $q$ values. The results also depend on the used scheme and range of box sizes, but again scheme B yields the best results. Using scheme B with box sizes from $L = 5$ to $19$ the scaling behavior can be well described by a power-law for $-4 \le q \le 3$ and $1 \le \alpha_q \le 2.8$ respectively.

\subsection{Three-dimensional wave functions}

\begin{figure}[b!]
    \centering
    \includegraphics[width=\columnwidth]{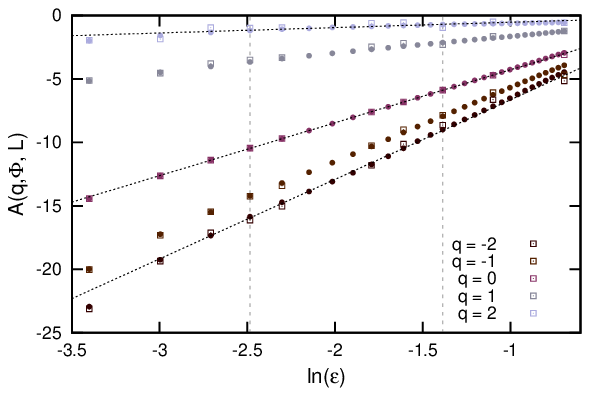}
    \includegraphics[width=\columnwidth]{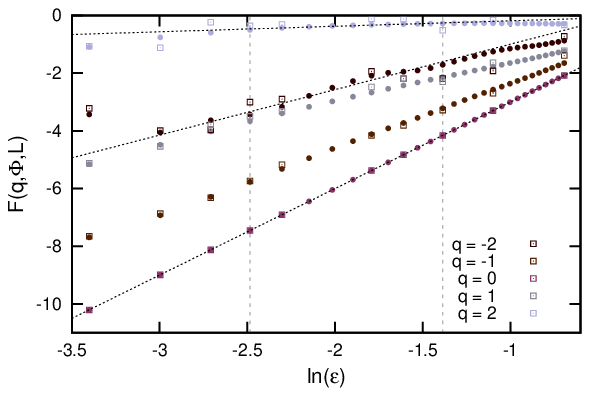}
    \caption{Same as Fig.\ \ref{fig:results-360-2D}, but for a wave function of the three-dimensional Anderson Hamiltonian for the linear system size $N = 60$ and the disorder strength $w = 16.5$. The least squares fits consider all box sizes in the range $5 \le L \le N/4$ marked by the two vertical dashed lines.}
    \label{fig:results-60-3D}
\end{figure}

\begin{figure}[b!]
    \centering
    \includegraphics[width=\columnwidth]{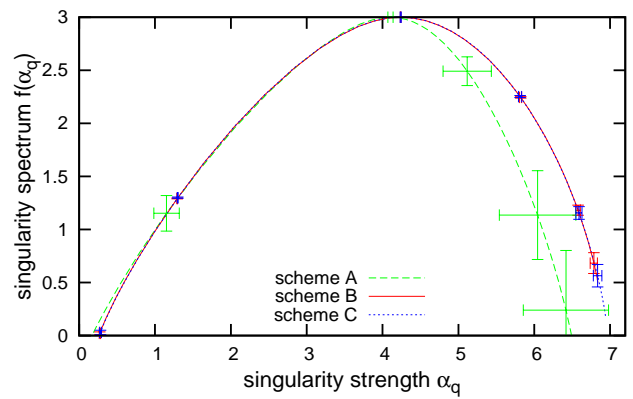}
    \caption{Singularity spectrum $f(\alpha_q)$ for the three different methods for the system in Fig.\ \ref{fig:results-60-3D} with considered box sizes from $L = 10$ to $N/2$. The plot shows $1\sigma$ error bars for integer $q$ from $-3$ to $3$.}
    \label{fig:singspec-60-3D}
\end{figure}

Finally, we have performed the same calculations also for the three-dimensional Anderson model for the disorder strength $w = 16.5$ at the mobility edge and a linear system size $N=60$. At the mobility edge in three dimensions the wave functions are known to be multifractals \cite{PhysRevLett.1991.Schreiber}. The results for the scaling of the quantities $A(q,\Phi,L)$ and $F(q,\Phi,L)$ are shown in Fig.\ \ref{fig:results-60-3D} and the corresponding singularity spectrum is shown in Fig.\ \ref{fig:singspec-60-3D}. Again both quantities, $A(q,\Phi,L)$ and $F(q,\Phi,L)$, show a power-law behavior and yield qualitatively similar results as in two dimensions for the partitioning schemes A and B. However, for the singularity spectrum we can observe a clear difference. While the results for the schemes B and C are nearly identical, the singularity spectrum obtained by the partitioning scheme A shows significant differences for negative values of $q$ (large $\alpha_q$). In Fig.\ \ref{fig:results-60-3D} we see that the average values of $A(q,\Phi,L)$ and $F(q,\Phi,L)$  are already quite different for negative values of $q$. This is probably related to the rather small system size. Hence, the distribution of the box probabilities is already rather different for relatively small box sizes. Further, due to the identical distribution of box probabilities for the partitioning schemes B and C, the only differences are the different number of considered data points for the least squares fit and hence the singularity spectra for both schemes are very similar.

\begin{figure}[t!]
    \centering
    \includegraphics[width=\columnwidth]{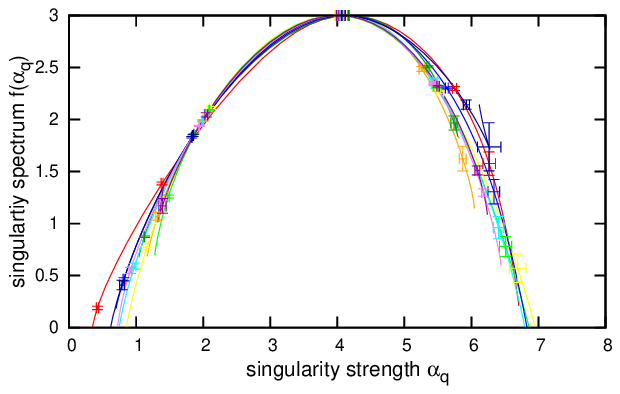}
    \includegraphics[width=\columnwidth]{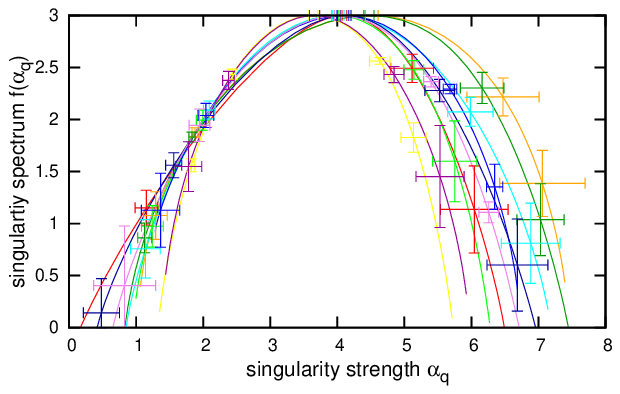}
    \caption{Singularity spectrum $f(\alpha_q)$ for wave functions at $E = 0$ for 10 different disorder realisations of the three-dimensional Anderson Hamiltonian for the system size $N = 60$ and the disorder strength $w = 16.5$ computed with scheme B with $5 \le L \le 15$ (top) and scheme A with $5 \le L \le 30$ (bottom). The plot shows 1$\sigma$ error bars for integer $q$ from $-2$ to $2$, and identical disorder realizations/wave functions are used for schemes A and B.}
    \label{fig:singspec-60-3D-all}
\end{figure}

\begin{figure}[t!]
    \centering
    \includegraphics[width=\columnwidth]{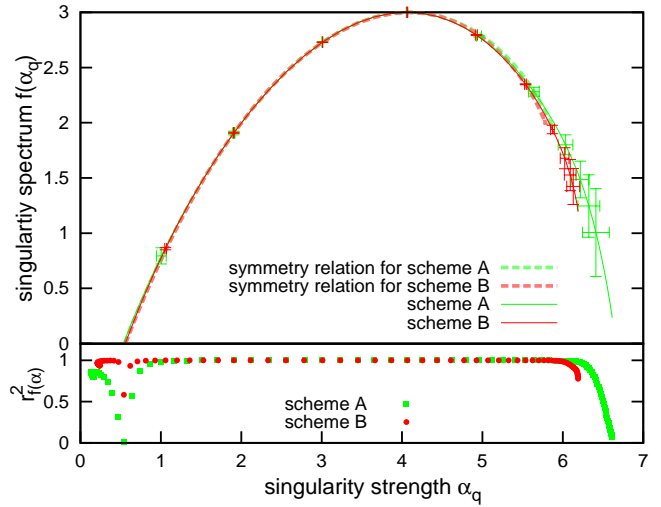}
    \caption{Ensemble average of the singularity spectrum (solid lines) and the correlation coefficients $r_{f(\alpha)}^2$ of the fractal dimensions for the wave functions in Fig.\ \ref{fig:singspec-60-3D-all}. The error bars reflect only the accuracy of the least squares fit for $q$ from $-1.5$ to $3$. We check whether the results fulfill the symmetry relation in Eq.\ \eqref{equ:symmetrie} by plotting the quantity $f(2d - \alpha_q) - d + \alpha_q$ (dashed lines).}
    \label{fig:singspec-60-3D-ens}
\end{figure}

All results show that scheme B significantly reduces the fit errors. This can be expected to ultimately lead to a better convergence of the ensemble average. To demonstrate this, we also show the singularity spectrum for wave functions closest to $E = 0$ for ten different disorder realizations and the corresponding fit errors for the schemes A and B in Fig.\ \ref{fig:singspec-60-3D-all}. For moderate values of $q$ scheme B shows very small fluctuations of the scaling exponents, which are in all cases smaller than for scheme A. We also present the corresponding ensemble average from these ten wave functions in Fig.\ \ref{fig:singspec-60-3D-ens}, which is obtained from \cite{EPJB.2009.Rodriguez}
\begin{align}
\label{equ:ensemble-average}
    \alpha_q &= \lim_{\varepsilon \to 0} \frac{1}{\ln \varepsilon}\frac{1}{\langle P_q \rangle} \bigg\langle \sum_{b=1}^{B} \mu_b^q(\Phi, L) \ln \mu_b(\Phi, L) \bigg\rangle \\
    f_q &= \lim_{\varepsilon \to 0} \frac{1}{\ln \varepsilon} \bigg[ \frac{q}{\langle P_q \rangle} \bigg\langle \sum_{b=1}^{B} \mu_b^q(\Phi, L) \ln \mu_b(\Phi, L) \bigg\rangle  - \ln \langle P_q \rangle \bigg]  \nonumber.
\end{align}
The error bars indicate the standard deviation due to the least squares fit and are not due to the fluctuations of the points for different disorder realizations. We also show the squares $r_{f(\alpha)}^2$ of the correlation coefficients of the fractal dimension for both methods. The standard deviations are quite small for both methods although we average only over ten wave functions. However, method B shows a significantly better performance for large positive and negative $q$.

\begin{figure}[t!]
    \centering
    \includegraphics[width=\columnwidth]{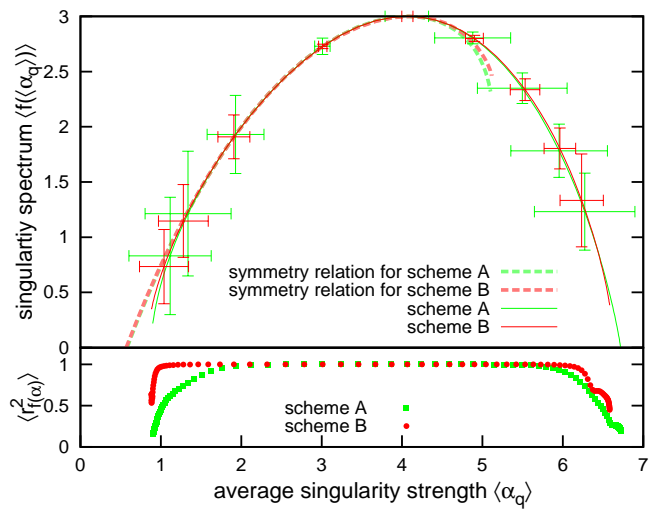}
    \caption{Typical average of the singularity spectra (solid lines) and the averaged correlation coefficients $\langle r_{f(\alpha)}^2 \rangle$ of the fractal dimensions for the wave functions in Fig.\ \ref{fig:singspec-60-3D-all}. The error bars reflect the standard deviation from the averaged exponents $\langle \alpha_q \rangle$ and $\langle f_q \rangle$ for $q$ from $-2$ to $2$. We again check whether the results fulfill the symmetry relation in Eq.\ \eqref{equ:symmetrie} (dashed lines).}
    \label{fig:singspec-60-3D-typ}
\end{figure}

We also like to point out that in literature different approaches for the determination of averages and errors in multifractal analysis can be found. On the one hand the exponents $\alpha_q$ and $f_q$ were determined from a least squares fit according to Eq.\ \eqref{equ:ensemble-average} \cite{EPJB.2009.Rodriguez,PhysRevB.2011.Rodriguez}, and on the other hand they were computed from the average of $\alpha_q(\Phi)$ and $f_q(\Phi)$ obtained for the single wave functions \cite{PhysRevB.1995.Grussbach,PhysRevB.1997.Milde}. The first method yields the ensemble average and the second method corresponds to the so-called typical average of the multifractal properties of the wave functions. For reasons of comparison we show in Fig.\ \ref{fig:singspec-60-3D-typ} the results for the typical average of the ten disorder realizations in Fig.\ \ref{fig:singspec-60-3D-all}. The results for scheme A and B are quite close but the error bars for scheme B are again significantly smaller. The main differences between the ensemble and the typical average occur for large positive and negative $q$ because for these cases the typical average is dominated by the behavior of a single wave function \cite{PhysRevB.2008.Vasquez}.

Further, in literature the error bars denote often different errors of the method like e.g.\ the standard deviation of the least squares fit \cite{PhysRevLett.1989.Chabra,ZPhysB.1991.Pook} or the standard deviation from averaging over the exponents obtained for different disorder realizations \cite{PhysRevB.1995.Grussbach,CompPhysComm.1999.Schreiber}. More advanced analyses use error propagation and consider correlations for the data points in the least squares fits according to Eq.\ \eqref{equ:ensemble-average} \cite{PhysRevB.2011.Rodriguez}.

In addition, we compare the resulting singularity spectrum of the ensemble average in Fig.\ \ref{fig:singspec-60-3D-ens} and of the typical average in Fig.\ \ref{fig:singspec-60-3D-typ} to the symmetry relation
\begin{equation}
    \label{equ:symmetrie}
    f(\alpha_q) = f(2d - \alpha_q) - d + \alpha_q
\end{equation}
proposed by Mirlin et al.\ \cite{PhysRevLett.2006.Mirlin}. We observe that this symmetry relation is fulfilled well for the ensemble and the typical average for positive $q$. However, for large $\alpha_q$ the ensemble averaged data show some differences from the symmetry relations, and we find considerable deviations for the typical average. Further, the proposed upper bound $\alpha_q < 2d$ is satisfied neither for the ensemble average nor for the typical average \cite{PhysRevLett.2006.Mirlin} although the results of scheme B for the ensemble average in Fig.\ \ref{fig:singspec-60-3D-ens} are quite close to this upper bound within the error margins. These singularity strengths correspond to large negative values of $q$ and, therefore, small errors are strongly enhanced in this region and the fit quality is decreasing (cp.\ also Fig.\ \ref{fig:singspec-360-2Db}). This has been found before in many other numerical studies. For instance, the results by Rodriguez et al.\ show that the bound $\alpha_q < 2d$ is approached for larger system sizes \cite{PhysRevB.2011.Rodriguez,PhysRevLett.2009.Rodriguez}. However, the comparison between the numerical results for scheme A with this symmetry relation shows bigger differences. The observations for the symmetry relation and for the ensemble average in Fig.\ \ref{fig:singspec-60-3D-ens} support the finding that scheme B can help to improve the ensemble average in the multifractal analysis.

\section{Conclusion}
\label{sec:conclusion}

We have compared different partitioning schemes in the box-counting algorithm in order to improve the evaluation of Eqs.\ \eqref{equ:A} and \eqref{equ:F} for a single wave function. We believe that the partitioning scheme considering unrestricted ratios $1/\varepsilon$ of the system size and the box size as well as an average over all box origins is better suited than the best method proposed by Rodriguez et al. \cite{EPJB.2009.Rodriguez}. This conclusion is based in particular on the observation that our method leads to smaller error bounds in the least squares fits and on the fact that it is not restricted to system sizes with preferable large numbers of factors. Our scheme is not meant to improve the ensemble average procedure and therefore it does not interfere with the established error estimates of this average. We  improve the evaluation of Eqs. \eqref{equ:A} and \eqref{equ:F} by avoiding any bias in the choice of the origin of the boxes and by consequently averaging in Eq. \eqref{equ:average} over all possible choices. We found that the average over different box origins yields a better estimate of the distribution for the box probability already for moderately large box sizes $L$. Due to the thus improved values of $A(q,\Phi,L)$ or $F(q,\Phi,L)$, a better fit of the singularity strength $\alpha_q$ and the fractal dimension $f_q$ becomes feasible. Further, we showed that only small box sizes should be considered for the fit. Additionally, our scheme allows us to consider many more box sizes $L$, what also helps to improve the fits of $\alpha_q$ and $f_q$.

Rodriguez et al. also studied the multifractal properties of the Anderson Hamiltonian with system-size scaling although it is computationally more demanding \cite{PhysRevB.2008.Rodriguez3,PhysRevB.2008.Vasquez}. They found that system-size scaling yields better fit results than the box-scaling approach. However, they compared their results with that of partitioning scheme A. Hence, a comparison with the better partitioning scheme B is necessary to validate the better performance of the system-size scaling approach. But we expect that also for the system-size scaling scheme B is superior in comparison with scheme A, not only because of the averaging over the box origins, which yields more accurate values of $A(q,\Phi,L)$ or $F(q,\Phi,L)$, but also because there is no restriction for the system sizes: in A again one needs a value of $N$ which factorizes into many factors while in B every $N$ can be included in the evaluation.

\newcommand{\noopsort}[1]{} \newcommand{\printfirst}[2]{#1}
\newcommand{\singleletter}[1]{#1} \newcommand{\switchargs}[2]{#2#1}

\end{document}